\documentclass[aos, preprint]{imsart}

\usepackage{amsmath, amssymb, amsthm}
\usepackage{bbold}
\usepackage{mathtools}
\usepackage{color,xcolor}
\usepackage{graphicx}
\usepackage{subfig}
\usepackage{cite}
\usepackage{enumerate} 
\usepackage[normalem]{ulem}
\usepackage{cancel} 

\usepackage[margin=1.0in]{geometry}
\usepackage{parskip} 

\RequirePackage[colorlinks,citecolor=blue,urlcolor=blue]{hyperref}

\newcommand{\abs}[1]{\left|#1\right|}

\newcommand{\A}[1]{\textbf{A}_{#1}}
\newcommand{\B}[1]{\textbf{B}_{#1}}
\newcommand{\C}[1]{\textbf{C}_{#1}}

\newcommand{\OS}[1]{\mathcal{O}^*(#1)}

\DeclareMathOperator{\E}{E}

\DeclareMathOperator{\Prob}{\text{Pr}}
\DeclareMathOperator{\ind}{\perp \!\!\! \perp }

\newcommand{\BB}[1]{\bar{\textbf{B}}_{#1}}

\DeclarePairedDelimiter{\ceil}{\lceil}{\rceil}
\newtheorem{theorem}{Theorem}
\newtheorem{theorem*}{Theorem}
\newtheorem{lemma}[theorem]{Lemma}
\newtheorem{corollary}{Corollary}[theorem]

\newtheorem*{assumption*}{\assumptionnumber}
\providecommand{\assumptionnumber}{}
\makeatletter
\newenvironment{assumption}[1]
 {%
  \renewcommand{\assumptionnumber}{AS#1}%
  \begin{assumption*}%
  \protected@edef\@currentlabel{#1}%
 }
 {%
  \end{assumption*}
 }
\makeatother
\newcommand{\asref}[1]{AS\ref{#1}}

\begin{document}

\begin{frontmatter}

\title{Finite Sample Complexity of Sequential Monte Carlo Estimators}
\runtitle{ Finite Sample Complexity of SMC}

\author{\fnms{Joe} \snm{Marion}\corref{}\thanksref{t1}\thanksref{t2}
\ead[label=e1]{jcm98@stat.duke.edu}}
\author{\fnms{Joseph} 
\snm{Mathews}
\ead[label=e2]{joseph.mathews@duke.edu}}
\and
\author{\fnms{Scott C.} \snm{Schmidler}\thanksref{t1}\thanksref{t3}
\ead[label=e3]{schmidler@stat.duke.edu}}

\thankstext{t1}{Supported by the National Science Foundation (NSF) grant DMS-1638521 to the Statistical and Applied Mathematical Sciences Institute} 
\thankstext{t2}{Supported by NSF research traineeship grant DMS-1045153}
\thankstext{t3}{Supported by NSF grant DMS-1407622}
\affiliation{Duke University}

\begin{abstract}
We present bounds for the finite sample error of sequential Monte Carlo samplers on static spaces.  Our approach explicitly relates the performance of the algorithm to properties of the chosen sequence of distributions and mixing properties of the associated Markov kernels.  This allows us to give the first finite sample comparison to other Monte Carlo schemes. We obtain bounds for the complexity of sequential Monte Carlo approximations for a variety of target distributions including finite spaces, product measures, and log-concave distributions including Bayesian logistic regression. The bounds obtained are within a logarithmic factor of similar bounds obtainable for Markov chain Monte Carlo.
\end{abstract}


\end{frontmatter}

\section{Introduction}
Sequential Monte Carlo samplers (SMC)~\cite{chopin2002sequential,del2006sequential} have recently received attention as an alternative to Markov chain Monte Carlo (MCMC) for Bayesian inference problems. Practitioners cite a variety of reasons for using SMC over MCMC. One reason is that it provides a natural estimate of the normalizing constant and may be the preferred method for estimating marginal likelihoods or Bayes factors~\cite{neal2001annealed, zhou2016toward, cerou2012}. SMC algorithms are well-suited for parallel computing environments and have been shown to provide large improvements in performance relative to non-parallel algorithms~\cite{lee2010utility, Durham2014}. A variety of methods have been developed to facilitate the implementation of SMC on graphics processing units or clusters of computers~\cite{lee2010utility,lee2016forest, Verge2015, Durham2014}. Finally, SMC exhibits similar properties to tempering~\cite{neal2001annealed,chopin2002sequential,del2006sequential, Durham2014}, making it well suited for difficult or multimodal problems. While these properties could make SMC a competitive alternative to MCMC, they have rarely been verified theoretically. 

The preponderance of SMC theory focuses on the asymptotic regime, where the number of particles approaches infinity. The existence of a central limit theorem for the SMC estimator was established by Del Moral and Guionnet~\cite{delmoral1999} and extended by Chopin~\cite{chopin2004central}. A similar CLT was shown to hold for adaptive resampling methods by Douc and Moulines~\cite{douc2008} and later by Beskos et al.~\cite{beskos2016}. Other asymptotic theory includes the work of Jasra et al.~\cite{jasra2015error}, who proved a bound on the asymptotic variance under local mixing assumptions. Beskos et al.~\cite{beskos2014}, showed non-degeneracy of the particle approximation as the dimension increases for problems with product measures. Eberle and Marinelli~\cite{eberle2013quantitative,Eberle_convergenceof} developed asymptotic error bounds for the continuous time analogue of SMC. Other relevant theoretical results in the particle filtering literature include the uniform convergence results of Del Moral and Miclo~\cite{delmoral2007} and Crisan~\cite{Crisan2001}, the Hillbert metric stability results of Le Gland and Oudjane~\cite{legland2004}, and the CLT of Künsch~\cite{kunsch2005}.

Finite sample results have been largely concerned with the $L_p$ stability of SMC. This includes  Whiteley~\cite{whiteley2012sequential}, who developed $L_p$ error bounds on non-compact spaces using drift and minorization conditions; C{\'e}rou et al.~\cite{cerou2011}, who provided finite sample bounds on the $L_2$ relative error of the particle system; and Schweizer~\cite{schweizer2012non} who demonstrated $L_p$ stability for finite-sample SMC on compact spaces using global and local mixing conditions. While these finite sample results are useful for establishing general characteristics of SMC, they depend on expectations and norms of the associated Feynman-Kac measures, making them difficult to evaluate in practice.

In this paper we develop finite sample bounds which enable the characterization of SMC as a randomized approximation scheme. Let $\pi$ be a target measure on $\mathcal{X}$ and $f:\mathcal{X}\rightarrow \mathcal{R}$ a bounded measurable function. Our main result provides, for any error tolerance $\epsilon>0$ and error probability $\delta\in (0,1/4]$, a choice of the number of particles $N$ and the number of Markov chain transitions $t$ at each step of the algorithm to ensure 
$$\Pr(|\hat{f} -\pi(f)|<\epsilon) \geq 1-\delta$$
where $\pi(f)$ denotes the expectation of $f$ with respect to $\pi$ and $\hat f$ is the SMC estimator. In contrast to other finite sample SMC bounds, we make explicit the dependence of $N$ and $t$ on properties of the distribution sequence (an upper bound to the weights, an upper bound on the ratio of normalizing constants between adjacent interpolating distributions, the mixing times of the Markov kernels) and the specified $\epsilon$ and $\delta$. The primary advantage of such bounds is that they allow for the interrogation of the algorithm, identifying how changes in the construction of the distribution sequence and choice of Markov kernels affect the computational cost of the estimator. The bound provided here also facilitates explicit comparison with other methods such as MCMC, potentially identifying situations where one or the other method may be preferred.  Our approach differs from previous analyses by focusing on the marginal distribution of individual particles rather than following the Feynman-Kac semi-group approach popularized by Del Moral~\cite{DelMoral2004Feynman}. We use an inductive approach to controlling the error at each step of the algorithm, developing sufficient conditions for propagating forward accurate particle approximations with high probability.

The paper is structured as follows. Section~\ref{Sec:SMC} introduces some notation and describes the general form of the SMC algorithm studied in this paper. Section 3 provides a statement of our main result, an error bound for the SMC estimator. Section~\ref{Sec:ErrorBounds} presents the proof of our error bound, developing conditions for inductively controlling the error via a coupling argument. Section~\ref{Sec:GeometricMixtures} uses our bound to compare the performance of SMC with MCMC on sequences of distributions obtained  via geometric mixtures with application to finite state spaces. This comparison highlights important differences between the algorithms and provides some guidance on how to select the interpolating distributions. Section~\ref{sec:product_measures} uses our bounds to explore the scaling of the SMC with dimension on product measures, and compares our results to those obtained previously in asymptotic and continuous time settings. This example also demonstrates the utility of our bounds in comparing SMC behavior under distinct choices of distribution sequences, showing that when the target is Gaussian with precision $\phi$ a careful choice of intermediate distributions can decrease the complexity from exponential in $\phi$ to logarithmic. Section~\ref{sec:log_concave} considers the case of log-concave target distributions and provides an application to Bayesian logistic regression. To the best of our knowledge this represents the first non-asymptotic SMC bound on a problem of direct interest to Bayesian statistical practice.

\section{Sequential Monte Carlo}
\label{Sec:SMC}
Let $\pi$ be a target probability measure on a space $\mathcal{X}$ with $\sigma$-algebra $\mathcal{B}$ and dominating measure $\lambda(dx)$.  Consider a test function $f:\mathcal{X}\rightarrow\mathcal{R}$. Our goal is to quantify the finite sample error arising from estimating $\pi(f)$ using sequential Monte Carlo. In this section, we introduce our probabilistic setting and the SMC algorithm studied in this paper.

\subsection{Notation}\label{sec:notation}
Let $\mathcal{P}$ be the set of probability measures on $\mathcal{X}$ that are absolutely continuous with respect to $\lambda$ and $\mathcal{F}$ the set of measurable functions $f:\mathcal{X} \rightarrow\mathcal{R}$. Each measure acts on functions $f\in\mathcal{F}$ from the left by $\mu (f) = \int f(x)\mu(dx)$. A  measure $\nu\in\mathcal{P}$ is said to be {\it $\omega$-warm} with respect to $\mu$ for $\omega\geq1$ if $\omega =  \sup_{B\in\mathcal{B}} \;\nu(B) / \mu(B)$~\cite{Lovasz04, Vempala05}. Let $\mathcal{P}_\omega (\mu)$ be the set of all such measures.

Let $K:\mathcal{X}\times\mathcal{B}\rightarrow[0,1]$ be an ergodic Markov kernel with limiting distribution $\mu$.  Markov kernels operate on functions from the left $K f(x)  = \int K(x, dy) f(y)$ and probability distributions from the right\break $\mu K(dx) = \int \mu(dy) K(y, dx)$. Define the mixing time of $K$ from an $\omega$-warm initial distribution by
$$
\tau(\epsilon, \omega) = \min\; \big\{t : \sup_{\nu \in \mathcal{P}_\omega(\mu)} \|\nu K^t - \mu\|_{\text{TV}}\leq \epsilon\big\}
$$
where $\|\cdot \|_{\text{TV}}$ is the total variation norm. Note that this is a somewhat weaker notion of mixing time than commonly used. In particular, obtaining samples from $\mu$ in polynomial time by simulating $K$ requires not only that $\tau_K$ grows at most polynomially in $1/\epsilon$ and $\omega$, but also the ability to draw an initial state from an $\omega$-warm distribution. Part of our result will be to show that SMC with appropriately chosen parameters guarantees an $\omega$-warm starting distribution.

When $K$ is irreducible, aperiodic, and $\mu$-invariant with spectral gap $\rho\in(0,1)$, for any $\nu << \mu$ and any positive integer $t$ we have:
\begin{equation*}\label{eq:spectral_gaps}
        %
        \frac{1}{2}\|\nu K^t - \mu\|_{\text{TV}} \leq
        \|\nu K^t - \mu\|_{2,\mu} 
        %
        \leq \|\nu - \mu\|_{2,\mu}\cdot (1-\rho)^t\\
        %
\end{equation*}
where $\|\nu - \mu\|_{2,\mu}^{2} = \int \left(\frac{\nu(dx)}{\mu(dx)}-1\right)^2\mu(dx)$ is the $\chi^2$  distance. Since the warmness condition $\nu \in \mathcal{P}_\omega (\mu)$ implies $\|\nu - \mu\|_{2,\mu} \leq \omega-1$, this yields the mixing time bound  $\tau_K(\epsilon,\omega) \leq \frac{1}{\rho}\left(\log(2\epsilon^{-1}) + \log(\omega-1)\right)$.

\subsection{Sequential Monte-Carlo}\label{sec:algorithm_definition}
In sequential Monte Carlo a collection of particles transition through a sequence of measures $\mu_0,...,\mu_S \in \mathcal{P}$ where $\mu_S = \pi$. Denote the density of each intermediate measure by $q_s(x)/z_s$.  In addition to the sequence of measures, we are given a collection of $\mu_s$ invariant Markov transition kernels $K_1,...,K_S$. 

In this paper we consider the following sequential Monte Carlo algorithm. Initialize by drawing $N$ independent samples $X_0^{1:N} = \big( X_{0}^{1},...,X_{0}^{N} \big)$ from $\mu_0$. The realizations of these particles are denoted by $x_{0}^{1:N} = \big( x_{0}^{1},...,x_{0}^{N} \big)$. For $s=1,...,S$ perform the following:
\begin{enumerate}
    \item[(i)]  Assign each particle an importance sampling weight equal to the unnormalized density ratio:
    $$w_{s}\big(x_{s-1}^{n}\big) = \frac{q_s\big(x_{s-1}^{n}\big)}{q_{s-1}\big(x_{s-1}^{n}\big)}.$$
    \item[(ii)] Sample a new set of particles with replacement according to the weights (multinomial resampling):
    $$ \Pr\Big(\tilde{X}_{s}^{n} = x \;\big|\; X_{s-1}^{1:N} = x_{s-1}^{1:N}\Big) \propto \sum_{n=1}^N w_{s}\big(x_{s-1}^{n}\big) \cdot \delta_{x_{s-1}^{n}}(x).$$
    \item[(iii)] Apply $t$ steps of the kernel $K_{s}$ to each re-sampled particle, producing $X_{s}^{1:N}$: 
    $$ \Pr\left(X_{s}^{n} \in dx \;\mid\; \tilde{X}_{s}^{n} =\tilde x_{s}^{n}\right) = K^t\big(\tilde x_{s}^{n}, \;dx \;\big).$$
\end{enumerate}
The average weight at each step is $\hat{w}_{s} = N^{-1}\sum_{n=1}^N w_{s}\big(x^{n}_{s-1}\big)$. When the algorithm is finished, the SMC estimate of $\pi(f)$ is $\hat{f} = \frac{1}{N} \sum_{n=1}^N f\big(x_{S}^{n}\big)$. Intuitively, the weighting step identifies particles in regions of high relative density, while the resampling step oversamples particles in under-represented regions while removing particles with low weights, so that computation is not wasted on particles in low density areas. This comes at the cost of increased dependence among the particles, often referred to as particle degeneracy, and characterized by multiple particles sharing same value immediately after resampling. The last step combats this degeneracy by evolving the resampled particles under the Markov kernel. This contracts the marginal distribution $\tilde{\mu}_s$ towards the desired distribution $\mu_s$ and reduces dependence between the particles.

\subsection{Probability space}
\begin{figure}[t]
\centering
\includegraphics[width=0.9\textwidth]{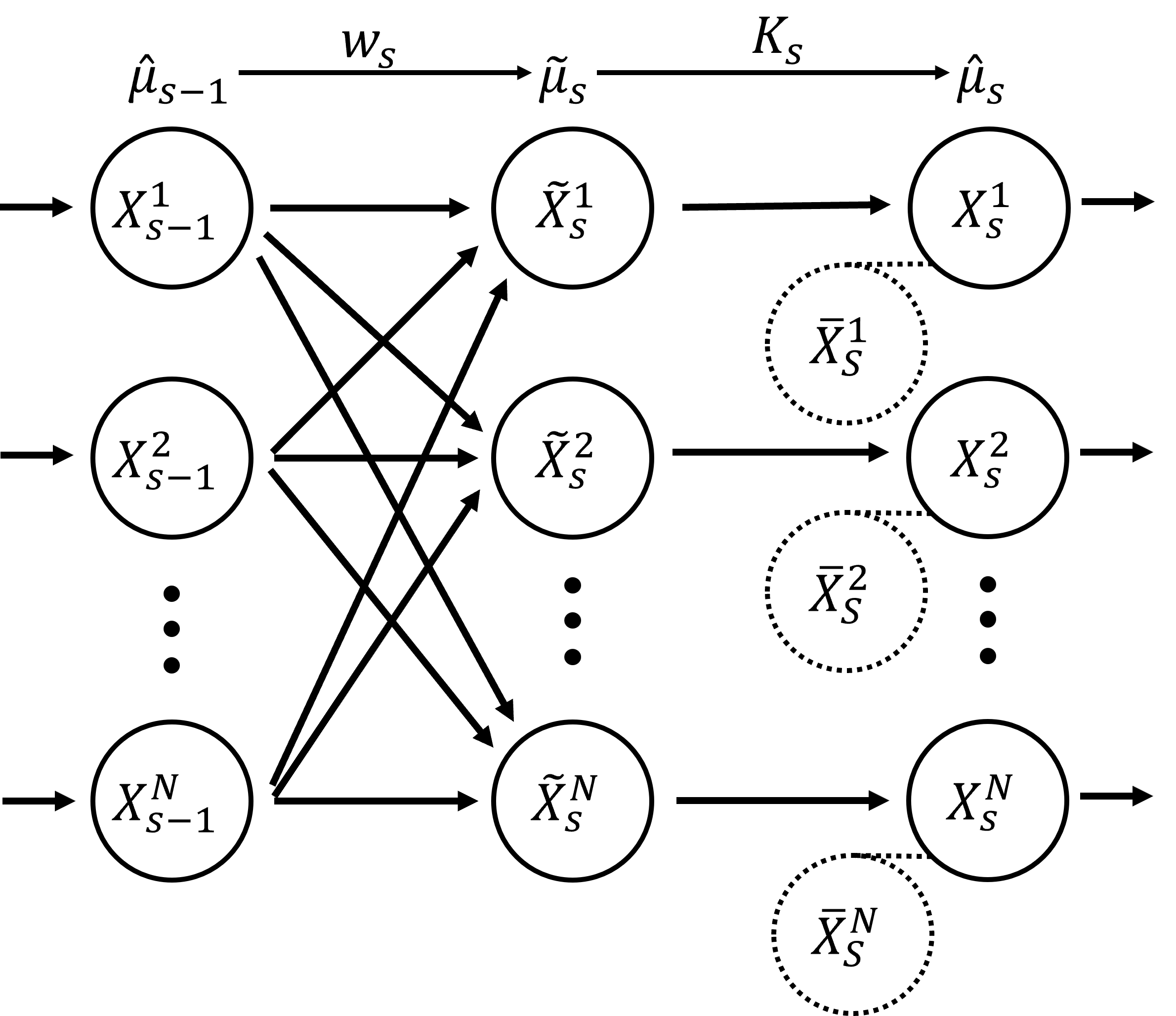}
\caption{Marginal distributions and dependence structure of the particles.}
\label{fig:dependence}
\end{figure}

This section contains a brief description of the probability space of the particle system; a full construction can be found in~\cite{DelMoral2004Feynman} chapter 3. For $1 \leq n \leq N$ and $0 \leq s \leq S$ let $X_s^{1:n} = (X_s^{1},...,X_s^n)$. At each step of the algorithm the particles evolve according to the following non-homogeneous Markov chain with law:

\begin{align*}
       \Pr\big(X_0^{1:N} \in dx^{1:N}) &= \prod_{n=1}^N \mu_0(dx^n) \\ 
       \Pr\big(\tilde{X}_s^{1:N} \in dx^{1:N} \mid X_{s-1}^{1:N}) &= \prod_{n=1}^N\sum_{m=1}^N \frac{w_s\big(X_{s\-1}^m\big)}{\hat{w}_s} \cdot \delta_{X_{s\-1}^m}(dx^n) \\ 
       \Pr\big(X_s^{1:N} \in dx^{1:N} \mid \tilde{X}_{s}^{1:N}) &= \prod_{n=1}^N K_s^t(\tilde{X}_s^n, dx^n) 
\end{align*}
where $dx^{1:N} = d(x^1,...,x^N)$ and $\delta_X(dx)$ denotes the Dirac measure. These three equations define the joint distribution of the random variables $X_0^{1:N},  X_1^{1:N}, \allowbreak \tilde{X}_1^{1:N}, \ldots , \tilde{X}_S^{1:N}, X_S^{1:N}$ associated with the particle system. By symmetry, the marginal distributions of $\tilde X_s^i$ are identical for all $i$ and denoted by $\tilde\mu_{s}$; similarly, denote the marginal distribution of the  $X_s^i$'s by $\hat\mu_{s}=\tilde{\mu}_s K_s^t$. Thus at the beginning of each step $s$ of the SMC algorithm, the particles $X_{s-1}^{1:N}$ are identically distributed according to $\hat\mu_{s-1}$. After resampling (step (ii)), the particles remain identically distributed and have marginal distribution $\tilde X_{s}^{i}\sim \tilde\mu_{s}$. Applying $t$ steps of $K_s$ (step (iii)) to each particle changes the marginal distribution to $\hat \mu_s = \tilde \mu_s K_s^t$. The dependence structure and marginal distributions of the particles are displayed visually by the probabilistic graphical model in Figure~\ref{fig:dependence}.

Our approach to controlling the error of SMC depends on relating the marginal distributions of the particles to the pre-specified interpolating distributions $\mu_{s}$. At each step of the algorithm, we show that the marginal distributions $\hat\mu_s$ remain close to the desired distribution $\mu_s$.

\subsection{Coupled random variables}\label{sec:coupling}
Our proof technique introduces additional random variables $\bar{X}_1^{n},\ldots, \bar{X}_S^{n}$ on $\left(\mathcal{X}, \mathcal{B}, \lambda\right)$ with each $\bar{X}_s^n$ having marginal distribution {\it exactly} $\mu_s$. These $\bar{X}_s^n$'s, which are constructed using a maximal coupling approach (see Appendix~\ref{app:coupling}), are represented in Figure~\ref{fig:dependence} by  dashed lines, indicating that they exist only as a theoretical construction rather than a direct output of the algorithm.  Inclusion of these variables expands the probability space under consideration to the set of random variables $(X^{1:N}_{0:S}, \tilde{X}^{1:N}_{1:S},\bar{X}^{1:N}_{1:S})$ jointly defined on $\mathcal{X}^{N(S+1)} \times \mathcal{X}^{NS} \times \mathcal{X}^{NS}$ (with corresponding $\sigma$-algebra $\mathcal{B}^{N(3S+1)}$).
Throughout the paper, marginalization and expectation are defined with respect to this joint distribution unless otherwise specified. 

\section{Main result}
We now state the main result of the paper, which is proven in Section~\ref{Sec:ErrorBounds}. This result bounds the probability of error of the SMC estimator as a function of $N$ and $t$, and allows us to establish SMC as a {\it randomized approximation scheme}, i.e. an algorithm which guarantees $|\hat{f} - \pi(f)|<\epsilon$ with probability at least $1-\delta$ (see e.g.~\cite{Motwani1995RandomizedA}). It is standard to show this bound holds with probability at least $3/4$; this can then be improved to probability $1-\delta$ by running the algorithm $\mathcal{O}\big(\log(1/\delta)\big)$ times and taking the median of the estimates (see Lemma 6.1 of \cite{JERRUM1986}). The main result uses the following assumptions.

\begin{assumption}{1}\label{as:1}
    Let $\sup w_s\left(x\right) \leq W$ and $ z_{s-1}/z_s \leq Z$ so that for $s=1,\ldots,S$: $$\frac{q_s(x)/z_s}{q_{s-1}(x)/z _{s-1}} \leq W\cdot Z$$
\end{assumption}
\begin{assumption}{2}\label{as:2}
    \parbox{\linewidth}{$K_s$ has limiting distribution $\mu_s$ with mixing time $\tau_s(\epsilon, \omega)$.}
\end{assumption}
\begin{theorem}
\label{thm:total_bound}
Assume~\asref{as:1} and \asref{as:2}. Fix $\epsilon > 0$ and sample $X_{0}^{1:N}$ independently from $\mu_0$. Let
\begin{enumerate}
    \item $N\geq \frac{1}{2}\log\big(128S\big)\cdot \max\big\{9W^2 Z^2, \frac{1}{\epsilon^2}\big\}.$
    \item $t \geq \max_{s}\; \tau_s \big( \frac{1}{8NS},\; 2\Big).$
\end{enumerate}
Then for any $f\in\mathcal{F}$ with $|f|\leq 1$:
$$\big|\hat{f} - \pi(f)\big| \leq \epsilon$$
with probability at least $3/4$.
\end{theorem}

When each $K_s$ is geometrically ergodic (has a spectral gap), \asref{as:2} can be replaced by the following assumption:
\begin{assumption}{3}\label{as:3}
    $K_s$ is irreducible, aperiodic, and $\mu_s$-invariant with spectral gap $\rho_s$ and $0<\rho_s \leq \rho \leq 1$.
\end{assumption}
This provides the following corollary to Theorem~\ref{thm:total_bound}:
\begin{corollary}\label{cor:spectral_bound}
Assume~\asref{as:1} and \asref{as:3}. Fix $\epsilon > 0$ and assume $X_{0}^{1:N}$ are sampled independently from $\mu_0$. Let
\begin{enumerate}
    \item $N\geq \frac{1}{2}\log\big(128S\big)\cdot \max\big\{9W^2 Z^2, \frac{1}{\epsilon^2}\big\}.$
    \item $t \geq \log\left(16NS\right) / \rho.$
\end{enumerate}
Then for any $f\in\mathcal{F}$ with $|f|\leq 1$:
$$\big|\hat{f} - \pi(f)\big| \leq \epsilon$$
with probability at least $3/4$.
\end{corollary}

\section{Error bounds}
\label{Sec:ErrorBounds}
The proof of Theorem~\ref{thm:total_bound} uses a coupling argument to bound the error of the SMC estimator. Our approach is inspired by that of Lovasz and Vempala (2006)~\cite{LovaszON3}, who construct a set of independent couplings for non-homogeneous (annealed) Markov processes. However, a key difference is that the resampling step in SMC introduces dependency between the particle trajectories. Here, we exploit conditional independence properties of the particle system to couple the marginal distributions of the particle trajectories, inductively, in order to establish the error bounds for Theorem~\ref{thm:total_bound}. At the end of this section, we show that our construction implies a coupling for the joint distribution of the particles at each step.

Our approach establishes conditions under which the SMC particles at each step $s$ are coupled with high probability to a set of particles drawn exactly from the target distribution $\mu_s$, given the previous step. An inductive argument then establishes that these conditions hold marginally  at each $s$ (and therefore for last step $S$ in particular) with high probability. The resulting coupling is used to establish concentration of sample means of the original particles around their target expectations.

Let $X_s^{1:N}$ be the particles produced at step $s$ of the process.  
We will construct for each $s$ a set of random variables $\bar{X}_s^{1:N}$ satisfying $\bar{X}_s^n \sim \mu_s$  marginally and such that $\Pr(X_s^n = \bar{X}_s^n) \geq 1-\delta$. To begin the induction, at step $s=0$ this is satisfied trivially by taking $\bar{X}_0^i = X_0^i$ since $X_0^{1:N}$ are drawn independently from $\mu_0$.  We will show that this can be done for all $s$; the construction of $\bar{X}_s^{1:N}$ is given in Lemma~\ref{lemma:coupling}.  For steps $s=0,...,S-1$, define the events:
\begin{align*}
    \A{s} & = \Big{\{}X_s^{1:N} = \bar{X}_s^{1:N}\Big{\}} .\\
    \B{s} & = \Big{\{}|\hat{w}_{s+1} - \mu_s(w_{s+1})| \leq \mu_s(w_{s+1})/3\Big{\}}.\\
    \C{s} & = \A{s} \cap \B{s}.
\end{align*}
Event $\textbf{A}_s$ represents the coupling event between the SMC particles and particles drawn directly from the target distribution.  $\textbf{B}_s$ is the event that the empirical estimator $\hat{w}_s$ lies within a relative interval around its mean $\mu_{s}(w_{s+1}) = z_{s+1}/z_{s}$.  We will show that, by choosing $N$ and $t$ as appropriate functions of $\tau_s$, $W$, and $Z$, the SMC algorithm ensures that
$\Pr(\textbf{C}_{S-1})$ is sufficiently large to construct a high probability coupling at step $S$. The final step of this induction allows us to bound the error of the final particle approximation $\hat{f}$ with high probability.

\subsection{Inductive argument and coupling construction}
Our inductive step consists of two pieces:
\begin{enumerate}
    \item Show that $\Pr\left(\A{s}\right) \geq (1-\delta) \cdot \Pr\left(\C{s-1}\right).$
    \item Show that $\Pr \left( \C{s} \right) \geq \left(1-\delta\right)\cdot \Pr\left (\C{s-1} \right)- \delta'.$
\end{enumerate}
The terms $\delta,\delta' \in (0, 1)$ are error probabilities associated with the number of samples $N$ and number of Markov kernel transitions $t$, respectively. First, we show (1), that coupling occurs at step $s$ with high relative probability.
\begin{lemma}\label{lemma:coupling}
Assume~\asref{as:2}. Suppose $\Pr\left(\C{s-1} \right) \geq \frac{3}{2 \omega}$ for some $\omega > \frac{3}{2}$. Then for any $0<\delta<1$ and $t \geq \tau_{s}\big(\frac{\delta}{N}, \omega\big)$:
\begin{equation*}
    \Pr\left(\A{s}\right) \geq (1-\delta) \cdot \Pr\left(\C{s-1}\right) \geq (1-\delta)\cdot \frac{3}{2\omega}.
\end{equation*}
\end{lemma}
To prove Lemma~\ref{lemma:coupling}, we will need the following additional lemma:

\begin{lemma}\label{lemma:warmness}
The marginal distribution $\tilde{\mu}_s$ of the resampled particles $\tilde{X}_s^n$  conditional on $\textbf{C}_{s-1}$ is $\omega$-warm with respect to $\mu_{s}$. 
\end{lemma}
\begin{proof}
Let $B \subset \mathcal{X}$ be a measurable set and let $\tilde{\mu}_{s}\left(dx \mid \C{s-1}\right)$ be the conditional distribution.
\begin{align*}
 \tilde{\mu}_{s}\left(B \mid \C{s-1}\right) = & \Pr\left(\tilde{X}^{n}_{s} \in B \;|\; \textbf{C}_{s-1} \right) \nonumber\\
        = &\sum^{N}_{j=1}\E\left[\frac{w_{s}(X^{j}_{s-1})}{\sum^{N}_{k=1}w_{s}(X^{k}_{s-1})}\delta\left(X^{j}_{s-1} \in B\right)  \;\bigg|\; \  \textbf{C}_{s-1}\right] \\
        = &\sum^{N}_{j=1}\E\left[\frac{w_{s}(\bar{X}^{j}_{s-1})}{\sum^{N}_{k=1}w_{s}(\bar{X}^{k}_{s-1})}\delta\left(\bar{X}^{j}_{s-1} \in B\right)  \;\bigg|\; \  \textbf{C}_{s-1}\right] \\
        \leq &\frac{3}{2N}\sum^{N}_{j=1}\E\left[\frac{w_{s}(\bar{X}^{j}_{s-1})}{\mu_{s-1}(w_{s})}\delta\left(\bar{X}^{j}_{s-1} \in B\right) \;\bigg|\; \textbf{C}_{s-1} \right] \\
        \leq &\frac{3}{2N} \sum^{N}_{j=1}\E\left[ \frac{\mu_{s}(\bar{X}^{j}_{s-1})}{\mu_{s-1}(\bar{X}^{j}_{s-1})}\delta\left(\bar{X}^{j}_{s-1} \in B\right) \right] \frac{1}{\Pr\left(\textbf{C}_{s-1} \right)} \\
         \leq &\omega \cdot \mu_{s}(B).
\end{align*}
The third and fourth lines follow since $\{X^{1:N}_{s-1} = \bar{X}^{1:N}_{s-1} \} \cap \{ \frac{2}{3}\mu_{s-1}(w_{s}) \leq \hat{w}_{s} \}  \supset \textbf{C}_{s-1}$, and the final line follows since $\Prob(\textbf{C}_{s-1}) \geq \frac{3}{2 \omega}$ by assumption. Thus we have $\tilde{\mu}_{s}\left(dx \mid \C{s-1}\right) \in \mathcal{P}_\omega (\mu_{s})$.
\end{proof}

With Lemma~\ref{lemma:warmness} in hand, we now proceed with the proof of Lemma~\ref{lemma:coupling}:

\begin{proof}
The coupling probability can be lower bounded as follows:
\begin{equation*}
    \begin{split}
        \Pr\left(\A{s}\right) &\geq  \Pr\left(\A{s} \cap \C{s-1}\right) \\
        &=  \Pr\left(\A{s} \;\mid\; \C{s-1}\right) \cdot \Pr\left(\C{s-1}\right).\\
    \end{split}
\end{equation*}
The result follows by showing that $\Pr\left(\A{s} \;\mid\; \C{s-1}\right) \geq 1-\delta$.  To do so, we note that Lemma~\ref{lemma:warmness} ensures that
each resampled particle $\tilde{X}_s^n$ has marginal distribution $\tilde{\mu}_{s}\left(dx \mid \C{s-1}\right)$ that is $\omega$-warm with respect to $\mu_s$.  After resampling, $t$ steps of $K_{s}$ are applied to each sample $\tilde{X}_s^n$ independently to obtain $X_{s}^{1:N}$. ~\asref{as:2} and the choice of $t$ thus ensures:
$$||\hat{\mu}_{s}\left(dx \mid \C{s-1}\right) - \mu_s(dx)||_{\text{TV}} \leq \frac{\delta}{N}.$$ 
For each particle we construct a coupled particle $\bar{X}_s^n$ via a maximal coupling construction (see e.g.~\cite{roberts2004general}) as described in Section~\ref{sec:coupling}. This results in a coupled pair of random variables $(X_s^n,\bar{X}_s^n)$, where  $\bar{X}_s^n \sim \mu_s$ and $X_s^n$ is the particle simulated by the SMC algorithm (so $X_s^n \sim K_s(\tilde{X}_s^n,\cdot)$). The bound on the total variation distance lower bounds the probability of coupling when condition $\C{s-1}$ holds:
$$\Pr\left( X_{s}^n = \bar{X}_s^n \;\mid\; \C{s-1} \right) \geq 1-  \frac{\delta}{N}.$$ 
This construction is carried out independently for each particle. Taking an intersection bound over the particles gives $\Pr\left(\A{s} \;\mid\; \C{s-1}\right) \geq 1-\delta$.\end{proof}
Lemma~\ref{lemma:coupling} establishes part (1) of our inductive step. To show part (2), we must show that $\B{s}$ also occurs with high probability given $\C{s-1}$.  To do so, we will a establish concentration property of the particle estimator. We begin with Lemma~\ref{lemma:independent}, which establishes the independence of the constructed $\bar{X}^{1:N}_{s}$.
\begin{lemma}\label{lemma:independent}
$\bar{X}^{1:N}_{s} \overset{iid}{\sim} \mu_{s}$ for each $s = 1,\ldots,S$.
\end{lemma}
\begin{proof} 
Let $B_{s} = B_{1} \times \ldots \times B_{N} \subset \mathcal{X}^{N}$ be an arbitrary measurable set. Let $E_{n} = \{\bar{X}^{n}_{s} \in B_{n}\}$. Using the fact $\bar{X}_s^k \ind \bar{X}_s^{[-k]} \mid \tilde{X}^k_s$ (see Lemma \ref{cond_ind_property} in Appendix~\ref{app:coupling}),
\begin{align*}
   \Prob(\cap^{N}_{n=1}E_{n}) &=  \E\left[\Prob(\cap^{N}_{n=1}E_{n} \mid \tilde{X}^{1:N}_{s}) \right] \\
    &= \E\left[\prod \Prob(E_{k} \mid  \tilde{X}^{k}_{s} )\right] \\
    &= \E\left[\prod \Prob(E_{k}) \right]= \prod \Prob(E_{k}).
\end{align*}
The third equality follows since $\bar{X}^{k}_{s} \ind \tilde{X}^{1:N}_{s}$ (see Lemma \ref{lemma:marginal_independent} in Appendix~\ref{app:coupling}).
\end{proof}

The next lemma establishes a concentration property of the particle system.  Let $g \in \mathcal{F}$ be a bounded function with $|g|\leq G$. Define $\hat{g} = \frac{1}{N}\sum_{n=1}^N g(X_s^n)$ to be the SMC estimator of $\mu_s(g)$ at step $s$. Since we have $X_s^{1:N}=\bar{X}_s^{1:N}$ with high probability (Lemma~\ref{lemma:coupling}) and the $\bar{X}_s^n$'s are independent  (Lemma~\ref{lemma:independent}), we have that $\hat{g}$ concentrates around $\mu_s(g)$ with high probability:

\begin{lemma}\label{lemma:concentration}
Assume~\asref{as:2}. Suppose $\Pr\left(\C{s-1} \right) \geq \frac{3}{2 \omega}$ for some $\omega > \frac{3}{2}$. For any $g \in \mathcal{F}$ with $|g| \leq G$, let $\hat{g} = \frac{1}{N} \sum_{n=1} ^N g(X_s^n)$.  Fix $0<\delta<1$, $0<\delta'<1$, and $\epsilon>0$. Then for any $N \geq \frac{G^2}{2\epsilon^2}\log(2/\delta')$ and $t \geq \tau_{s}\big(\frac{\delta}{N}, \omega\big)$,
\begin{equation*}
    \Pr \left( |\hat{g} - \mu_s(g)| < \epsilon \right) \geq \left(1-\delta\right)\cdot \Pr\left (\C{s-1} \right)- \delta'.
\end{equation*}
\end{lemma}
\begin{proof}
Let $\bar{g} = \frac{1}{N}\sum_{i=1}^N g(\bar{X}_s^n)$  The choice of $N$ and Lemma~\ref{lemma:independent} give $\Pr\left( \BB{s} \right) \geq 1-\delta'$ $\Pr\left( |\bar{g}_{s} - \mu_s(g)| \geq \epsilon\right) \leq \delta'$ by H\"oeffding's inequality. Then:

\begin{align*}
\Pr\left(\abs{\hat{g} - \mu_s(g)} \geq \epsilon\right) & = 
\Pr\left( \left\{\abs{\hat{g} - \mu_s(g)} \geq \epsilon\right\} , \A{s}\right)+\Pr\left( \left\{\abs{\hat{g} - \mu_s(g)} \geq \epsilon\right\} , \A{s}^c\right)\\
& \leq \Pr\left( \left\{\abs{\hat{g} - \mu_s(g)} \geq \epsilon\right\} ,  \A{s}\right)+\Pr\left(\A{s}^c\right) \\
& = \Pr\left( \left\{|\bar{g}_{s} - \mu_s(g)| \geq \epsilon\right\} , \A{s} \right)+\Pr\left(\A{s}^c\right) \\
& \leq \Pr\left( |\bar{g}_{s} - \mu_s(g)| \geq \epsilon \right)+\Pr\left(\A{s}^c\right)\\
& \leq \delta' + (1 - (1-\delta)\cdot\Pr(\C{s-1}))
\end{align*}
where the fifth line uses Lemma~\ref{lemma:coupling} (which requires~\asref{as:2}). The result follows.
\end{proof}
A similar proof with $g=w_{s+1}$, $G=W$, and $\epsilon= 1/3Z \leq \mu_s(w_{s+1}) / 3$ establishes part (2) of our inductive step:
\begin{corollary}
\label{cor:iteration}
Assume~\asref{as:1} and \asref{as:2}. Suppose $\Pr\left(\C{s-1} \right) \geq \frac{3}{2 \omega}$ for some $\omega > \frac{3}{2}$. Fix $0 < \delta < 1$ and $0 < \delta' < 1$. Then for any $N \geq \frac{9 W^2 Z^2}{2}\log(2/\delta')$ and $t \geq \tau_{s}\big(\frac{\delta}{N}, \omega\big)$:
\begin{equation*}
    \Pr \left( \C{s} \right) \geq \left(1-\delta\right)\cdot \Pr\left (\C{s-1} \right)- \delta'.
\end{equation*}
\end{corollary}
\begin{proof}
Let $\bar{w}_{s+1} = \frac{1}{N}\sum_{i=1}^N w_{s+1}(\bar{X}_s^n)$ and define:
\begin{equation*}
    \BB{s} = \Big{\{}|\bar{w}_{s+1} - \mu_s(w_{s+1})| < \epsilon\Big{\}}
\end{equation*}
the event that the weights of the coupled particles concentrate around their mean $\mu_s(w_{s+1})$. ~\asref{as:1},~\asref{as:2} the choice of $N$ and Lemma~\ref{lemma:independent} give $\Pr\left( \BB{s}^C\right) \leq \delta'$ by H\"oeffding's inequality.

\begin{equation*}
    \begin{split}
    \Pr \left( \C{s}^C \right)&=\Pr\left( \B{s}^C \cap \A{s} \right) + \Pr\left(\A{s}^C \right)\\
    &=\Pr\left( \BB{s}^C \cap \A{s} \right) + \Pr\left(\A{s}^C \right)\\
    &\leq \Pr\left( \BB{s}^C \right) + \Pr\left(\A{s}^C \right)\\
    &\leq \delta' + (1 - (1-\delta)\cdot\Pr(\C{s-1}))\\
    \end{split}
\end{equation*}
where the fourth line uses Lemma~\ref{lemma:coupling} (which requires~\asref{as:2}) \end{proof}

The proof of Theorem~\ref{thm:total_bound} is completed by applying Corollary~\ref{cor:iteration} inductively to establish that $\A{S}$ holds with high probability.  The error of the final SMC estimator $\hat{f} = \frac{1}{N} \sum_{n=1}^N f\big(x_{S}^{n}\big)$ can then be controlled using Lemma~\ref{lemma:concentration}. 

\begin{proof}[Proof of Theorem 1]

By Lemma~\ref{lemma:concentration} the error of the SMC estimator $\hat{f}$ satisfies:
\begin{equation*}
    \begin{split}
        \Pr\left( |\hat{f} - \pi(f)|\leq \epsilon\right)     &\geq \left(1-\delta\right)\cdot \Pr\left (\C{S-1} \right)- \delta'.
    \end{split}
\end{equation*}
$\Pr\left (\C{S-1} \right)$ can be lower bounded by induction using Corollary~\ref{cor:iteration}. The base case is established by noting that $\Pr\left(\C{0}\right) \geq 1 - \delta'$  since $\A{0}$ holds by definition and $\B{0}$ follows from H\"oeffding's inequality. Repeated application of Corollary~\ref{cor:iteration} gives:
\begin{align*}
    \Pr\left(\C{s}\right) & \geq \left(1-\delta\right)^s\cdot(1-\delta') - \delta' \sum_{r=0}^{s-1}\left(1-\delta\right)^r \\
    &= \left(1-\delta\right)^s - \delta' \sum_{r=0}^{s}\left(1-\delta\right)^r \\
    &= \left(1-\delta\right)^s - \delta'\frac{1-\left(1-\delta\right)^{s+1}}{\delta}\\  
    &\geq \left(1-\delta\right)^s - \frac{\delta'}{\delta}.\\
\end{align*}
Selecting $\delta = 1/8S$ and $\delta' = 1/64S$ gives:
\begin{equation*}
    \begin{split}
        \Pr\left( |\hat{f} - \pi(f)| \leq \epsilon \right)  &\geq  \left(1-\delta\right)^S - \frac{\delta'}{\delta}\\
        & = \left(1 - 1/8S \right)^S  - \frac{1}{8}\\
        %
        & \geq 1 - 1/8 - 1/8\\
        &= 3/4.\\
    \end{split}
\end{equation*}
Theorem~\ref{thm:total_bound} follows by selecting $\omega=2$, $N\geq \frac{1}{2}\log\big(128S\big)\cdot\max\big\{9W^2 Z^2, \frac{1}{\epsilon^2}\big\}$ and $t \geq \max_{s}\; \tau_s \big( \frac{1}{8NS},\; 2\big)$.
\end{proof}
This proves the main result. Corollary~\ref{cor:spectral_bound} follows immediately using standard bound on the warm mixing time stated in section~\ref{sec:notation}. The requirement that $X_0^{1:N}$ are iid according to $\mu_0$ can be relaxed as long as $\textbf{C}_0$ holds with high probability. This might be the case, for example, when the initial particles are drawn using a rapidly mixing Markov chain.  In addition, Theorem~\ref{thm:total_bound} could be refined so that the $t$ and $N$ are allowed to depend on $s$, so long as the requirements of Lemma~\ref{lemma:coupling} and Corollary~\ref{cor:iteration} are satisfied. This modification would provide a more efficient bound, however, we've omitted this complication to ease the presentation.

In the remaining sections of the paper, we use the bounds provided by Theorem~\ref{thm:total_bound} and Corollary~\ref{cor:spectral_bound} to compare the complexity of SMC with that of MCMC in a variety of settings.  Before doing so, we conclude this section with a final result that follows from the proof of Theorem~\ref{thm:total_bound} given in this section.

\subsection{Approximate independence of particles}

While not necessary to prove Theorem~\ref{thm:total_bound}, the following corollary provides additional insight into the behavior of the particle system:
\begin{lemma}
\label{lemma:extra}
Assume~\asref{as:1} and \asref{as:2}. $X_s^{1:N}$ are approximately i.i.d. with distribution $\mu_s$; that is,
$$\|\mathcal{L}(X_s^{1:N} ) - \mu_s^N \|_{\text{TV}} \leq \frac{\delta_0}{2}.$$
\end{lemma}
\begin{proof}
By Lemma~\ref{lemma:warmness} and choice of $t=\tau_s(\frac{\delta_0}{2N},2)$ we have
$$\|\mathcal{L}(X_s^{n} ) - \mu_s \|_{\text{TV}}  = \|\hat{\mu}_s -
\mu_s\|_{\text{TV}} \leq \frac{\delta_0}{2N} \quad \text{ for all } n.$$
The coupling construction given in the previous section defines random
variables $\bar{X}_s^{n}$ for every $i$ such that 
$\bar{X}_s^{n} \sim \mu_s$ and $\text{Pr}(X_s^{n} \neq
\bar{X}_s^{n}) \leq \frac{\delta_0}{2N}$.  We then have
$\text{Pr}(X_s^{1:N} \neq \bar{X}_s^{1:N}) \leq
\frac{\delta_0}{2}$ by the union bound. 
But then $(X_s^{1:N}, \bar{X}_s^{1:N})$ form a coupling and by
the coupling inequality:
$$\|\mathcal{L}(X_s^{1:N} ) - \mathcal{L}(\bar{X}_s^{1:N})
\|_{\text{TV}} \leq \text{Pr}(X_s^{1:N} \neq \bar{X}_s^{1:N}).$$
By Lemma~\ref{lemma:independent} we have $\mathcal{L}(\bar{X}_s^{1:N}) =
\mu_s^{\otimes N}$ which establishes the result.
\end{proof}

This result tells us that Theorem~\ref{thm:total_bound} effectively describes a coupling of the full joint distribution of the particle system to the the target distribution $\mu_s^N$ at each step; that is, the particles are approximately iid $\mu_s$ for all $s$.

\section{SMC with geometric mixtures}
\label{Sec:GeometricMixtures}
Geometric mixtures are a common and straightforward way of specifying a sequence of SMC distributions. Consider the problem of sampling from $\pi$ having density $q_{\pi}(x)/z_{\pi}$ known up to $z_\pi$. Suppose we can efficiently draw independent samples from an initial distribution $\nu$ with density $q_{\nu}(x)/z_{\nu}$. 
Define the geometric mixture distribution $\mu_{\beta}$ for $\beta\in[0,1]$ by the unnormalized density
\begin{equation*}
    q_\beta(x) = q_{\pi}(x)^{\beta}\cdot q_{\nu}(x)^{(1-\beta)}.
\end{equation*}
As $\beta$ varies from 0 to 1, the distributions $q_\beta$ interpolate from initial distribution $\nu$ to the target distribution $\pi$. If $\nu$ is uniform, the $q_\beta$ are called {\it tempered} versions of $\pi$ and $\beta$ is called the inverse temperature. (In Bayesian statistical inference for posterior distribution $\pi$, $\nu$ is  often chosen instead to be the prior distribution~\cite{Durham2014,chopin2002sequential,zhou2016toward}).

A distribution sequence for SMC can be defined by evaluating $q_\beta$ at a finite set of $\beta$ values  $0=\beta_0,\beta_1,\ldots,\beta_S=1$. To simplify notation we index the mixture distributions by $s$ with $q_s(x) = q_{\beta_s}(x)$ and denote the normalizing constant $z_{\beta_s} = z_s$.  When the uniform distribution is improper or difficult to sample and no better choice is available, $\nu$ may be chosen as $q_\beta$ for some $1 > \beta >0$ (sufficiently high temperature) which is accessible via MCMC. Choosing the initial distribution to be either uniform or tempered is analogous to simulated annealing, starting from a relatively diffuse distribution and moving towards a more concentrated distribution of interest.

We consider the computational complexity of SMC using geometric mixtures, measured in terms of the number of total Markov kernel transitions $SNt$ required to obtain a $(\delta, \epsilon)$ randomized approximation scheme.  This serves as a measure of overall computational complexity, since the Markov kernel transitions tend to dominate the computational cost of the SMC algorithm. If parallel computing resources are available, the performance of SMC may be improved by a constant factor via parallelization, but the overall complexity of the bounds does not change. We note that the parallelization of SMC is not trivial due to the resampling step and that specific SMC algorithms have been developed for this computational approach~\cite{lee2016forest,Verge2015}.

\subsection{Finite sample bounds for SMC}
To specify the SMC algorithm, we need to choose a sequence of inverse temperatures $\beta_0,...,\beta_S$.  We assume the density ratio $\pi(x)/\nu(x)$ is bounded, so $\pi(x)/\nu(x) 
\leq \Gamma$, and choose $S=\ceil{\log \Gamma}$ and $\beta_s = s/S$. Let $\gamma = W\cdot Z$ where $W$ and $Z$ bound the maxima of $\sup w_s(x)$ and $z_{s-1}/z_s$ as defined in section~\ref{sec:algorithm_definition}; so $\gamma$ bounds the density ratio $\frac{q_s(x)/z_{s}}{q_{s-1}(x)/z_{s-1}}$ for all $s$. We will assume that for each $\beta\in(0,1]$ we can construct an ergodic Markov kernel $K_\beta$ with spectral gap $\rho_\beta$. 

Using this sequence of distributions we can apply Theorem~\ref{thm:total_bound}.
\begin{corollary}\label{cor:complexity_smc}
Let $\rho = \min_s \rho_{\beta_s}$ and fix  $\epsilon>0$ . Then for any $f\in\mathcal{F}$ with $|f|\leq 1$, the number of Markov kernel transitions required to ensure $|\hat{f}-\pi(f)|\leq \epsilon$ with probability at least $3/4$ is bounded above by
$$
\mathcal{O}^*\Bigg(\frac{\epsilon^{-2} \vee \gamma^{2}}{\rho}\cdot \log \Gamma \cdot \log^2\log\Gamma \Bigg).
$$
\end{corollary}
The notation $\mathcal{O}^*$ indicates that lower order terms ($\log\log\log \Gamma$, $\log \gamma$ and $\log 1/\epsilon$) have been omitted for readability. The $\mathcal{O}\big(\frac{1}{\rho}\big)$ term is the number of Markov chain transitions required to ensure that the marginal distribution $\hat{\mu}_s$ is sufficiently close to $\mu_s$ for each $s$. The $\mathcal{O}\big(\epsilon^{-2} \vee \gamma^2\big)$ term represents the number of particles required to both estimate $\pi (f)$ with sufficient accuracy and ensure that $z_{s}/z_{s-1}$ is estimated with sufficient relative accuracy at each step of the algorithm (condition $\B{s}$). The final $\mathcal{O}\big(\log^2 \log \Gamma\big)$ term is the additional factor required to ensure that the iteration conditions hold throughout the steps of the algorithm.

The quantity $\epsilon^{-2} \vee \gamma^{2}$ provides some insight into the sources of SMC error. When high accuracy is not required ($\epsilon$ is large), a large number of particles may still be required (according to our bounds) to approximate $z_{s-1}/z_{s}$ with small relative error (Lemma~\ref{lemma:concentration}) and maintain the stability of the algorithm. When $\gamma$ is large for some $s$ and   an insufficient number of particles are used, this likely  manifests as particle degeneracy. This can be mitigated by  choosing $S$ sufficiently large to ensure  that $\gamma$ is $O(\epsilon^{-2})$. This is in accordance with SMC folklore that suggests large numbers of steps with modest numbers of particles are preferable. 
\subsection{Comparison with importance sampling}
It is also instructive to use our bound to quantify the advantages of SMC over standard importance sampling. When the ratio of normalizing constants $z_\pi/z_\nu$  is unknown, the importance sampling estimator is $\sum_{n=1}^N \frac{q_\pi(x_{n})}{q_{\nu}(x_{n})}f(x_n) \big/ \sum_{n=1}^N \frac{q_\pi(x_{n})}{q_{\nu}(x_{n})}$. To ensure that the absolute error of the estimator is less than $\epsilon$, both the numerator $\frac{1}{N}\sum_{n=1}^N \frac{q_\pi(x_{n})}{q_{\nu}(x_{n})}f(x_n)$ and its normalization $\frac{1}{N}\sum_{n=1}^N \frac{q_\pi(x_{n})}{q_{\nu}(x_{n})}$ must be accurately estimated. The numerator is relatively easy to estimate and requires $\mathcal{O}\big(W^2/\epsilon^2\big)$ samples (Hoeffding) where $W = \sup q_\pi(x)/q_\nu(x)$. On the other hand, the normalization must have small {\it relative} error compared to $\nu\left(q_\pi / q_\nu\right)=z_\pi/z_\eta=Z^{-1}$, which requires $\mathcal{O}\big(\Gamma^2/\epsilon^2\big)$ samples, with $\Gamma = WZ$. Comparing this with Corollary~\ref{cor:complexity_smc}, we see that while the complexity of importance sampling is quadratic in $\Gamma$, SMC depends only logarithmically on $\Gamma$, at the cost of a factor of $\mathcal{O}\big(1/\rho\big)$.  For many problems of interest $\Gamma$ may be exponentially large (e.g. in the dimension of the problem) and SMC can be expected to substantially outperform importance sampling.

\subsection{Comparison of SMC and MCMC} We compare the bound for SMC given in Corollary~\ref{cor:complexity_smc} with a corresponding bound for an MCMC approximation.  The MCMC approximation is created by repeating the following $N$ times independently: 
draw an initial point from $\nu$ and simulate $t'$ steps of $K_1$. Write $\bar{f}$ to denote the estimator constructed from the resulting samples.
\begin{corollary}\label{cor:complexity_mcmc}
Fix $\epsilon > 0$. Then for any function $f\in\mathcal{F}$ with $|f|\leq 1$, the number of Markov kernel transitions required to ensure $|\bar{f}-\pi(f)| \leq \epsilon$ with probability at least $3/4$ is bounded above by
$$
\mathcal{O}^*\Bigg(\frac{1}{\rho_1}\cdot \log \Gamma \cdot \frac{1}{\epsilon^2}\Bigg).
$$
\end{corollary}
\begin{proof}
By assumption $\nu\in \mathcal{P}_{\Gamma}(\pi)$, so choosing $t'= \mathcal{O}\Big(\frac{\log (\Gamma/\epsilon)}{\rho_1}\Big)$ ensures that $||\nu K_{1}^{t'} - \pi ||_\text{TV} \leq \epsilon/2$ and therefore $|\nu K_{1}^{t'}f - \pi f|\leq \epsilon/2$. Choosing $N=\mathcal{O}\big(\epsilon^{-2}\big)$ ensures that $|\bar\pi f - \nu K_{1}^{t'} f|\leq \epsilon/2$ with probability at least $3/4$ by H\"oeffding's inequality. The result follows from the triangle inequality. 
\end{proof} 

An alternative approach is to run a single Markov chain to near stationarity and then continue taking samples every $\mathcal{O}(\rho_{1}^{-1})$ steps to obtain a sequence of approximately independent samples~\cite{kannan1996sampling,kannan1997}. The complexity is then $\mathcal{O}^*\left(\rho_1^{-1}\cdot\max(\log(\Gamma),\epsilon^{-2})\right)$ versus $\mathcal{O}^*\left(\rho_1^{-1}\cdot\log(\Gamma)\cdot\epsilon^{-2}\right)$. 

To simplify comparison of the bounds obtained in Corollaries~\ref{cor:complexity_smc} and~\ref{cor:complexity_mcmc}, we will take $\epsilon= \mathcal{O}(\gamma^{-1})$, although as noted above the SMC bound will not decrease for larger $\epsilon$. We see that the bound for SMC requires an additional factor of $\mathcal{O}\big(\log^2\log \Gamma\big)$ to ensure the induction condition at each step. Note also that the complexity of MCMC depends only on $\rho_1$ rather than $\rho$.  Typically the construction of $q_\beta$ will ensure  $\rho_s > \rho_1$ for $s > 1$, but depending on the choice of $\nu$ this need not always hold. Finally, while both bounds depend on $\Gamma$, the SMC bound also depends on the maximum density ratio $\gamma$  between any pair of neighboring distributions. If one density ratio is much larger than  $\epsilon^{-1}$, a large increase in $N$ is required to control the error at that step.  Choosing the $\beta$'s so that the ratios are close to $\epsilon^{-1}$ and approximately equal provides the smallest upper bound. This agrees with heuristics for the selection of inverse temperatures found in the simulated tempering literature~\cite{park2007choosing}, which aim to space distributions so that the ratios of normalizing constants between adjacent distributions are approximately equal. $\gamma$ may be controlled by choosing $S$ sufficiently large.
\subsubsection{Example: finite spaces}
Let $\mathcal{X}$ be a finite space with $\pi(x)\propto q(x)$ and $0< q(x)\leq 1$. Let $\pi_0 = \min \pi(x)$ and let $x_0=\arg\min\pi(x)$ be a state at which this is attained. Let initial distribution $\nu(x) = \mathbb{1}_{x=x_0}$ assign mass one to $x_0$, yielding bound $\Gamma \leq \frac{1}{\pi_0}$.  The complexity of Markov chain Monte Carlo estimator is bounded above by:
$$\mathcal{O}^*\bigg(\frac{1}{\rho_1}\cdot\log\Big(\frac{1}{\pi_0}\Big)\cdot \frac{1}{\epsilon^2} \bigg).$$
A comparable bound can be obtained for SMC using our results. Let $\mu_0 \propto \pi^{\beta_0}$ with $\beta_0 = 1/\ceil{\log\frac{1}{\pi_0}}$; 
samples from $\mu_0$ can be drawn in $\mathcal{O}\big(\frac{1}{\rho_{\beta_0}}\big)$ time using independent Markov chains beginning at $x_0$. Set $S= \ceil{\log\frac{1}{\pi_0}}-1$ and choose $\mu_s \propto q(x)^{\beta_s}$ with $\beta_s = \frac{s+1}{\ceil{\log\frac{1}{\pi_0}}}$ giving $\gamma\leq e$. Applying  Corollary~\ref{cor:complexity_smc}, the complexity of SMC is bounded above by 
$$\mathcal{O}^*\bigg(\frac{\epsilon^{-2}}{\rho}\cdot\log\Big(\frac{1}{\pi_0}\Big)\cdot \log^2\log\Big(\frac{1}{\pi_0}\Big)\bigg).$$ 
\section{SMC on product measures}\label{sec:product_measures}
Product measures have previously been used to assess the dimension dependence of SMC~\cite{beskos2014,schweizer2012non,eberle2013quantitative}. Consider again the setup of initial distribution $\nu$ and target distribution $\pi$, with weight $w(x) =  q_\pi(x)/q_\nu(x)$, $\Gamma \geq \pi(x) / \nu(x)$ a bound on the density ratio, and $K$ a geometrically ergodic, $\pi$-reversible Markov kernel with spectral gap $\rho >0$.

Define product measures $\pi^d$ and $\nu^d$ on $\mathcal{X}^d$ with corresponding weight $w^d=\prod_{i=1}^d \frac{q_\pi(x_i)}{q_\nu(x_i)}$, and define the $\pi^d$-invariant product kernel $K^d=\prod_{i=1}^d K(x_i, dx_i)$. The spectral gap of $K^d$ is independent of dimension $d$ for a product kernel~\cite{eberle2013quantitative, schweizer2012non}, though the computational cost of each kernel transition increases linearly in $d$. Choosing a geometric mixture sequence with $S=\mathcal{O}(d)$ and linearly spaced $\beta_s$ ensures that $\gamma = \mathcal{O}(\Gamma)$~\cite{schweizer2012non}. 
Applying Theorem~\ref{thm:total_bound} bounds the computational complexity in terms of  dimension:
$$
\mathcal{O}(d^2 \log^2 d).$$
This improves upon the $\mathcal{O}(d^3)$ finite sample results of Schweizer~\cite{schweizer2012non} and Eberle and Marinelli~\cite{eberle2013quantitative}, though it falls short of the $\mathcal{O}(d^2)$ rate obtained by Beskos et al.~\cite{beskos2014} in the limit of infinite particles and dimensions; the latter result of course requires no cost to control the finite-sample approximation error.

We can apply these bounds to investigate the effect of inverse temperature selection on the computational complexity in the case of Gaussian product measures.
\subsection{Example: spherical Gaussian in \textit{d}-dimensions}\label{sec:gaus1}
Let $\pi$ be $d$-dimensional spherical Gaussian centered at the origin with precision $\phi>1$ and unnormalized pdf $q(x|\phi) = \exp\big(-\frac{1}{2}\phi\cdot x^T x \big)$ for $x\in\mathcal{R}^d$. Since many posterior distributions arising from Bayesian analyses are well approximated by normal distributions as the number of observations grows, this serves as a model for understanding the performance of SMC on well-behaved posteriors more generally.

Let $\nu$ be the $d$-dimensional standard normal distribution and construct interpolating distributions using geometric mixtures with $S=d$ and $\beta_s = s/d$. Then $\mu_s$ is also spherical normal, characterized by precision $\phi_s = 1 + \frac{s}{d}(\phi-1)$. For the specified temperature sequence, the largest density ratio occurs at the first step of the algorithm, with $\gamma = \sup \mu_1(x) / \mu_0(x) = z_0/z_1 = \big(1 + \frac{\phi-1}{d}\big)^{d/2} \leq \exp\big(\frac{\phi-1}{2}\big)$ with $\gamma \approx \exp\big(\frac{\phi-1}{2}\big)$ for $d$ large. Hence for $d$ sufficiently large, the overall complexity of SMC is bounded above by
$$\mathcal{O}^*\Big(d^2 
\cdot \max\{\exp(\phi), \epsilon^{-2}\}\log^2 d \Big)$$
where $S=d$, $t=\OS{d \log d }$ and $N = \OS{\max\{\exp(\phi),  \epsilon^{-2}\} \cdot\log d}$. Note that while the complexity in $d$ remains $\mathcal{O}^*(d^2\log^2 d)$, there is an exponential dependence on $\phi$.  This comes from the first step of the algorithm, where the initial distribution is very flat relative to the first interpolating distribution and $z_1/z_0$ becomes exponentially (in $\phi$) small, requiring many samples to estimate with low relative error. 

This problem may be addressed by selecting a better temperature ladder which ensures that $\mu_{s}$ is not too peaked relative to $\mu_{s-1}$ at any step. Indeed, for the same number of intermediate distributions $S$, choosing a log-linear spacing on the precision reduces the dependence on $\phi$ to polynomial. More precisely, taking $\phi_s = \phi^{\frac{s}{d}}$ gives $\gamma \leq \sqrt{\phi}$ and yields an SMC bound of
$$\mathcal{O}^*\bigg(d^2 
\cdot  \max\Big\{\phi, \frac{1}{\epsilon^{2}}\Big\}\log^2 d \bigg)$$
where $S=d$, $t=\OS{d \log d }$ and $N = \OS{\max\{\phi,  \epsilon^{-2}\} \cdot\log d}$. This  demonstrates the importance of the choice of  interpolating distributions. In fact, the dependence on $\phi$ can be further reduced to logarithmic by choosing $S = d\ceil{\log\phi}$ and $\beta_s = \exp(s/d) \wedge 1$. Under this choice $\gamma\leq e^{\frac{1}{2}}$ and the complexity is bounded above by
$$\mathcal{O}^*\bigg(d^2\cdot  \frac{\log\phi \cdot \log^2 d}{\epsilon^2}\bigg)$$
where $S=d\ceil{\log\phi}$, $t=\OS{d \log d }$ and $N = \OS{\epsilon^{-2}\cdot\log d}$. These results illustrate how the availability of finite sample bounds can enable the selection of better distribution sequences for SMC. This example in particular has important implications for well-behaved Bayesian inference problems, where $\phi$ large corresponds to posterior distributions that are highly concentrated. This indicates the importance of carefully selecting the temperature ladder especially when data sets are large. 

\section{Log-concave distributions}\label{sec:log_concave}
Log-concave target distributions arise in many settings of interest. Log-concave sampling problems have been well studied; examples in statistics include Bayesian analysis of regression and logistic regression problems with priors corresponding to convex penalties, such as the Bayesian ridge or LASSO priors. In this section we apply our bounds to these log-concave problems, incorporating key results from Wu et al.~\cite{WuMALA}.

Let $\pi(x)\propto q(x)$ be a distribution on $\mathcal{R}^d$.  We say that $q$ is strongly log-concave if $q^{1-\alpha}(x)\cdot q^\alpha(y) < q\big(\alpha x + (1-\alpha) y\big)$ for $x, y \in \mathcal{R}^d$ and $\alpha\in(0,1)$. To be able to use the results of~\cite{WuMALA}, we will assume further that $\log q$ is both $L$-smooth and $m$-strongly concave, i.e. that 
\begin{equation*}
        -\frac{L}{2}||x-y||_2^2 \leq \log \frac{q(x)}{q(y)} - \nabla \log q(x)^T(x-y) \leq -\frac{m}{2}||x-y||_2^2
\end{equation*}
for all $x,y\in\mathcal{R}^d$.  This implies  $-L||x-x^*||_2^2 \leq 2\log q(x) \leq -m||x-x^*||_2^2$, where $x^*$ is the mode of $\pi$. Let $\kappa = L/k$ denote the condition number of $\log q(x)$. Intuitively, $\kappa$ is a measure of the curvature of the density $q$ and is large e.g. when one dimension has a large range relative to the others. 

We consider sampling from such distributions using SMC.  We choose $\mu_0 =  N(x^*, 1/L)$ and use a tempered sequence of interpolating distributions. Choosing $S = \ceil{d\kappa}$ and $\beta_s = s/S$ gives $W = 1$ and $Z=\mathcal{O}(1)$~\cite{LovaszON3}. We also restrict $\mathcal{R}^d$ to a ball $B$ of radius $ 4\sqrt{d/m}$ centered at $x^*$. This restriction ensures that the ratio of normalizing constants is bounded in the first step; a similar restriction is made in~\cite{LovaszON3}. Since $\pi(B) \geq 1-\epsilon/2$ this assumption has minimal impact on the results of our analysis~\cite{Dwivedi2018}. 

For our Markov kernel we use the Metropolis-adjusted Langevin algorithm (MALA) kernel. Slightly larger bounds are immediately available for other kernels, e.g the \textit{ball walk} and the \textit{hit-and-run} walk~\cite{LovaszGeometry}. Wu et al.~\cite{WuMALA} show that the mixing time of MALA on log-concave problems is 
$\mathcal{O}^*(\kappa \sqrt{d})$ when starting from a
warm initial distribution. Tempering $q$ does not change the condition number, so the mixing time of $K_s$ is the same for all $s$. Plugging this mixing time into our SMC bounds gives a complexity of
$$ \mathcal{O}^*\big(d^{3/2} \kappa^2\cdot  \log^2 (d \kappa) \big)$$
where $S=\ceil{d\kappa}$, $t=\OS{\kappa \sqrt{d} \log( d \kappa)}$ and $N = \OS{\epsilon^{-2}\cdot\log(d\kappa)}$. This is larger than the $\mathcal{O}^*\big(d \kappa \cdot  \log^3 (d\kappa) \big)$ obtained by Lee et al.~\cite{leeMala}. Besides the $\log^2 d\kappa$ term, which is the penalty our bound pays to control the worst case error across each step, the SMC bound grows quadratically in $\kappa$ (actually $\kappa^2\log^2 \kappa$) whereas the MCMC bound grows as $\kappa \log^3 \kappa$. This increased complexity comes from the difficulty in constructing an optimal path for SMC: since the ratio $z_\nu / z_\pi$ is bounded above by $\kappa^{d/2}$, we suspect there exists a path of length $d\log \kappa$ which ensures $Z\leq e^\frac{1}{2}$~\cite{LovaszON3}.(In fact, recent work by the authors shows that $S=O(\sqrt{d}\log(dk))$ can be achieved under an assumption of $L_2$ bound on the density ratio). Such a path would reduce the dependence on $\kappa$ from $\kappa^2$ to $\kappa\log\kappa$ and eliminate this difference in the bounds.

\subsection{Example: Bayesian logistic regression} 
Consider fitting a logistic regression model to a binary observation vector $Y\in\{0,1\}^n$ and associated matrix of covariates $X\in \mathcal{R}^{n\times p}$, via Bayesian inference. The corresponding likelihood is given by:
$$
p(Y| X, \beta) \propto \exp\Big(Y^T X\beta - \sum_{i=1}^n \log\big(1 +e^{X_i^T \beta}\big)\Big).
$$
Assign prior $p_0(\beta) =  N\big(0, \frac{\alpha}{n}(X^TX)^{-1}\big)$  with the parameter $\alpha$ controlling the strength of the prior shrinkage toward zero. The resulting posterior distribution $q(\beta) \propto p_0(\beta)p(Y\mid X,\beta)$ is log-concave and satisfies the above assumptions of $L$-smoothness and $m$-strong concavity with $L\leq (n/4 + \alpha)\cdot \sigma_{\max} $ and $m \geq \alpha \cdot \sigma_{\min} $ for $\sigma_{\max}$ and $\sigma_{\min}$ the largest and smallest eigenvalues of $(X^TX)^{-1}/n$, respectively~\cite{Dwivedi2018}. Inserting into our bounds gives an upper bound on the complexity of sampling via SMC:
$$
\mathcal{O}^*\bigg( \Big(\frac{d n}{\alpha} \cdot \frac{\sigma_{\max}}{\sigma_{\min}}\Big)^{2} \cdot  \log^2 \Big(\frac{d n}{\alpha} \cdot \frac{\sigma_{\max}}{\sigma_{\min}}\Big) \cdot \max \Big\{1, \sqrt{\frac{n}{d\alpha} \cdot \frac{\sigma_{\max}}{\sigma_{\min}}}\Big\}\Big).
$$
This example demonstrates the utility of our approach for practical problems: we are unaware of any previous finite-sample error bounds for non-trivial problems in Bayesian statistics using SMC. The dependence of the bound on $\sigma_{\max}/\sigma_{\min}$ can be removed be improving the condition number via pre-conditioning (see~\cite{Dalalyan2017}). 

\section{Conclusion}
The finite-sample bounds on SMC error provided here enable rigorous analysis of the computational complexity of SMC sampling algorithms on static spaces.  As we have demonstrated, this allows for interesting comparisons between the efficiency of various SMC sampling algorithms, including the crucial dependence on the choice of interpolating distributions.  However, significant areas remain for potential improvement of these bounds and extensions in future work.

The SMC bounds presented in sections~\ref{Sec:GeometricMixtures},~\ref{sec:product_measures}, and~\ref{sec:log_concave} suffer additional logarithmic complexity in $\Gamma$, $d$, and $\log d\kappa$ respectively in comparison to MCMC. This arises from the requirement that the worst-case error is controlled across all steps (ensuring $\textbf{C}_s$ for all $s$). It has been suggested to us that it may be possible to remove this through use of Talagrand's generic chaining method, and we are exploring this approach. 

Another area of interest is target distributions exhibiting multimodality, where Markov kernels may have good local mixing behaviour, yet exhibit poor mixing globally (e.g. \cite{Woodard2009torpid,DougParallel}). Sequential Monte Carlo has been observed to perform well empirically for some of these target distributions. This also was demonstrated asymptotically by Jasra et al.~\cite{jasra2015error} for some problems studied by \cite{Woodard09rapid,Woodard2009torpid}.  The bounds presented in this paper require a global mixing condition and would require modification to show the advantage of SMC in this setting. Incorporating local mixing conditions into our methods along the lines of \cite{Woodard09rapid,schweizer2012non,jasra2015error} would allow us to obtain results more directly comparable to \cite{Woodard09rapid,Woodard2009torpid} and answer the interesting question of whether such beneficial behavior persists outside the asymptotic setting.

Finally, our approach is well suited to comparison of the many variations on SMC sampling algorithms, and could be extended to include adaptive SMC methods.
Adaptive methods can exhibit substantial performance gains in practice through adaptive selection of distributions and Markov kernels, but theoretical results for these methods to date are limited to adaptive resampling times~\cite{douc2008,beskos2016}. The techniques described in this paper may be well suited to demonstrating the stability and usefulness of more general adaptive methods.

\appendix
\section{Additional coupling results}\label{app:coupling}
We first give an explicit maximal coupling construction of $(X^{i}_{s}, \bar{X}^{i}_{s})$ and end this section with supporting lemmas needed to prove Lemma \ref{lemma:independent}. We note that the construction given here is similar to the the one given in Proposition 3(g) of \cite{roberts2004general}. \par For $x \in \mathcal{X}$, let $\rho_{x}(\cdot)$ be a dominating measure for $K^{t}_{s}(x,\cdot)$ and $\mu_{s}(\cdot)$ with corresponding densities
\begin{align*}
    \frac{dK^{t}_{s}}{d\rho_{x}} := f_{x} , \ \frac{d\mu_{s}}{d\rho_{x}} := g_{x}.
\end{align*}
Set $ h_{x} = \min\{g_{x},f_{x}\}$. The subscript denotes the implicit dependence on $\tilde{X}^{i}_{s}$. Let
 \begin{align*}
     a_{x} = \int_{\mathcal{X}}h_{x} d\rho_{x}, \ \  b_{x} = \int_{\mathcal{X}} (f_{x} - h_{x}) d\rho_{x}, \ \ c_{x} = \int_{\mathcal{X}} (g_{x} - h_{x}) d\rho_{x}.
 \end{align*}
We formalize the coupling construction via a `coupling map' $C: \mathcal{X} \times \mathcal{B} \times \mathcal{B} \rightarrow [0,1]$, which transitions between states by the following procedure. Given $x \in \mathcal{X}$,
\begin{enumerate}
    \item Independently draw $Z^{i}_{s}$, $U^{i}_{s}$, and $V^{i}_{s}$ according to their corresponding distributions with densities  $h_{x}/a_{x}$, $(f_{x} - h_{x})/b_{x}$, and $(g_{x} - h_{x})/c_{x}$, respectively.
    \item Draw $I^{i}_{s}$ independently such that $\text{Pr}(I^{i}_{s} = 1) = a_{x}$ and $\text{Pr}(I^{i}_{s} = 0) = 1 - a_{x}$.
    \item If $I^{i}_{s} = 1$, the new state is $(Z^{i}_{s},Z^{i}_{s})$; else, the new state is $(U^{i}_{s},V^{i}_{s})$.
\end{enumerate}
Intuitively, $C$ maps $\tilde{X}^{i}_{s}$ to a pair of random variables $(X^{i}_{s}, \bar{X}^{i}_{s})$ such that 
$\mathcal{L}(X^{i}_{s}) = \hat{\mu}_{s}$, $\mathcal{L}(\bar{X}^{i}_{s}) = \mu_{s}$, and $\Prob(X^{i}_{s} = \bar{X}^{i}_{s}) = ||\hat{\mu}_{s}(\cdot) - \mu_{s}(\cdot)||_{\text{TV}}$. To see this, note that
\begin{align*}
    C(x,B \times \mathcal{X}) &= \text{Pr}(U^{i}_{s} \in B \cap V^{i}_{s} \in \mathcal{X})(1 - a_{x}) + \text{Pr}(Z^{i}_{s} \in B \cap \mathcal{X})a_{x}, \\
    &= \text{Pr}(U^{i}_{s} \in B)\text{Pr}(V^{i}_{s} \in \mathcal{X})(1 - a_{x}) + \text{Pr}(Z^{i}_{s} \in B)a_{x} \\
    &= K^{t}_{s}(x,B).
\end{align*}
Similarly, $C(x,\mathcal{X} \times B) = \mu_{s}(B)$. Finally, by construction we have $\Prob(X^{i}_{s} = \bar{X}^{i}_{s} | \tilde{X}^{i}_{s}) = ||K^{t}_{s}(\tilde{X}^{i}_{s}, \cdot)- \mu_{s}(\cdot)||_{\text{TV}}$; integrating with respect to $\tilde{\mu}_s$ then gives 
$\Prob(X^{i}_{s} = \bar{X}^{i}_{s}) = ||\hat{\mu}_{s}(\cdot) - \mu_{s}(\cdot)||_{\text{TV}}$. 

The next two lemmas are used to prove Lemma \ref{lemma:independent}.
\begin{lemma}\label{cond_ind_property}
We have $\bar{X}^{k}_{s} \ind \bar{X}^{[-k]}_{s} \ | \ \tilde{X}^{k}_{s} \text{ for }k=1,\ldots,N$.
\end{lemma}
\begin{proof}
By construction, we have $ (X^{k}_{s}, \bar{X}^{k}_{s}) \sim  C(\tilde{X}^k_{s}, \cdot)$.
By definition, $(Z^{k}_{s}, U^{k}_{s}, V^{k}_{s})$ are independent across $k = 1,\ldots,N$ conditional on $\tilde{X}^{k}_{s}$. Since $C(\tilde{X}^{k}_{s}, \cdot)$ sets $(X^{k}_{s}, \bar{X}^{k}_{s})$ is equal to either $(Z^{k}_{s},Z^{k}_{s})$ or $(U^{k}_{s},V^{k}_{s})$, the result follows immediately.
\end{proof}
Lemma \ref{cond_ind_property} allows us to factor the joint distribution of the constructed particles into a product of conditional distributions. This is used in Lemma~\ref{lemma:independent}, along with the following Lemma, to establish marginal independence of the  $\bar{X}^{k}_{s}$.
\begin{lemma}\label{lemma:marginal_independent}
Suppose $ (X^{k}_{s}, \bar{X}^{k}_{s}) \overset{ind.}{\sim} C(\tilde{X}^{k}_{s}, \cdot)$ for $k=1,\ldots,N$. Then $\bar{X}^{k}_{s} \ind \tilde{X}^{k}_{s}$.
\end{lemma}
\begin{proof}
Let $B_{1}, B_{2} \subset \mathcal{X}$. We show $\text{Pr}(\bar{X}^{k}_{s} \in B_{1} \cap \tilde{X}^{k}_{s} \in B_{2}) = \text{Pr}(\bar{X}^{k}_{s} \in B_{1}) \text{Pr}(\tilde{X}^{k}_{s} \in B_{2})$. Suppose $\tilde{\mu}_{s}(B_{2}) > 0$, otherwise the result holds trivially. Notice
\begin{align*}
    \text{Pr}(\bar{X}^{k}_{s} \in B_{1} | \tilde{X}^{k}_{s} \in B_{2}) = \frac{1}{\tilde{\mu}_{s}(B_{2})}\int_{B_{2}}C(x, \mathcal{X} \times B_{1}) \tilde{\mu}_{s}(dx) =  \text{Pr}(\bar{X}^{k}_{s} \in B_{1}).
\end{align*}
Hence, $\text{Pr}(\bar{X}^{k}_{s} \in B_{1} \cap \tilde{X}^{k}_{s} \in B_{2}) =  \text{Pr}(\bar{X}^{k}_{s} \in B_{1})  \text{Pr}(\tilde{X}^{k}_{s} \in B_{2})$.
\end{proof}

\bibliographystyle{unsrt}
\bibliography{mybib.bib}

\end{document}